\documentclass[a4paper,11pt]{article}
\usepackage[11pt]{moresize}

\usepackage[a4paper, total={6in, 9in}]{geometry}
\usepackage{graphicx}
\usepackage{mathtools}
\usepackage{dcolumn}
\usepackage{float}
\usepackage{blindtext}
\usepackage{hyperref}
\usepackage{rotating}
\usepackage{booktabs}
\usepackage{array}
\usepackage{caption}
\usepackage[toc,page]{appendix}
\usepackage{appendix}
\usepackage{listings}
\usepackage{lscape}
\usepackage{multirow}
\usepackage{verbatim}
\usepackage[table]{xcolor}
\usepackage[colorinlistoftodos]{todonotes}
\usepackage{subcaption}
\usepackage{algorithmicx}
\usepackage[ruled,vlined]{algorithm2e}
\usepackage{algpseudocode}

\usepackage{graphicx}
\usepackage{lmodern,textcomp}
\usepackage[autostyle=true]{csquotes}

\definecolor{firebrick}{rgb}{0.69, 0.13, 0.13}
\definecolor{blue(pigment)}{rgb}{0.2, 0.2, 0.6}
\hypersetup{colorlinks=true, citecolor=blue(pigment), urlcolor  = cyan, linkcolor= firebrick, pdfborder = {0 0 1}}

\begin{document}

\title{\textbf{Developing a real estate yield investment device using granular data and machine learning}
\footnote{\textbf{\textit{Acknowledgments:}} We thank Diego Azqueta-Oyarzun and Guillermina Gavaldon Hernandez for valuable comments. We alone are responsible for any errors.}}

\author{
  Azqueta-Gavaldon, Monica\thanks{E-mail: \texttt{m.azquetag@gmail.com}} \\
  \textit{AstraZeneca Computational Pathology}
  \and
  Azqueta-Gavaldon, Gonzalo\thanks{E-mail: \texttt{gonzalo.azqueta@gmail.com }} \\
  \textit{University of Strathclyde}
  \and
  Azqueta-Gavaldon, Inigo\thanks{E-mail: \texttt{inigo.azqueta@gmail.com}} \\
  \textit{Technische Universität München}
  \and
  Azqueta-Gavaldon, Andres\thanks{E-mail: \texttt{a.azqueta-gavaldon.1@research.gla.ac.uk}} \\
    \textit{University of Glasgow}
}

\maketitle

\begin{abstract}
This project aims at creating an investment device to help investors determine which real estate units have a higher return to investment in Madrid. To do so, we gather data from \href{Idealista.com}{Idealista.com}, a real estate web-page with millions of real estate units across Spain, Italy and Portugal. In this preliminary version, we present the road map on how we gather the data; descriptive statistics of the 8,121 real estate units gathered  (rental and sale); build a return index based on the difference in prices of rental and sale units (per neighborhood and size) and introduce machine learning algorithms for rental real estate price prediction. 
\end{abstract}

\providecommand{\keywords}[1]{\textbf{\textit{keywords---}} #1}



\clearpage{}

\section{Introduction}

The rising uncertainties in the current and future economic outlook and low interest rates likely to remain so for the next couple of years, often leads to a low and risky return to investment. Moreover, rising inequalities often lead to rental prices to increase by more than sale prices, which offer a higher rental yield (the return earned when renting out a purchased property). Latest research shows that this yield ranges from 4.40\% to 5.15\% in Barcelona and Madrid and most importantly, it has been increasing during the last couple of years (\cite{Delmendo2020}). This preliminary work examines the rental yield of several real estate units in Madrid by making use of information gathered from Idealista.com and machine learning algorithms. \\

Real-state value is usually estimated by taking factors into account such as the size of a property, its location, pricing of similar neighboring properties etc. A would-be buyer or seller is thus influenced and restricted by the available real-state information that can be retrieved. Knowing the exact value of a property is paramount for all the parties involved in its transaction, and basing it on the available information can lead to biases. A seller can for instance dictate a property's value based on its size, amenities, number of rooms, bathrooms etc. However many latent factors (factors that are hard to take into account) influence the market value of the property such as views, neighborhood appeal, price by area etc. With large amounts of real-estate data and using statistical and novel machine learning algorithms, these latent features can be used to overcome biases and generate a more accurate picture of a property's value.\\ 

In Spain, there is a lower proportion of people living in rental houses compared to other EU members. However, in recent years there has been an increasing trend of living in rental houses rather than in owned property. More specifically, there has been an increase in medium to long term rental contracts, that is, non-tourist tenants, as the work of \cite{lopez2019recent} shows. This increase has been especially strong in cities like Barcelona and Madrid. The estimation of a property's rental price thus gains importance given this trend.\\

In our work, we first present how we obtain and prepare (merge and clean) the data from Spain`s biggest real estate portal; \href{Idealsita.com}{Idealsita.com}. We then offer an overview of the data and variables that we obtained while produce a rental yield index for each of neighborhood and property sizes in Madrid. To do so, we use the rental prices and the most likely mortgage payment for the sale units. In other words, we present a method to evaluate the profitability on different neighborhoods and property sizes in Madrid based on the average purchase and average rental prices. We find that the highest index can be found in the neighborhood of Opanel (south-west of Madrid city centre) across those units between 30 to 60 square meters. When it comes to bigger apartments, those between 60 to 90 square meters, we find that the neighborhood of Los Angeles (south-side of Madrid) display the higher yield. \\

We then present different machine learning algorithms trained to predict rental prices and evaluated them. As a benchmark we use multivariate linear regressions which can explain around 62\% of the variance of the dependent variable, rental price, by including only three variables; the size, whether or not the apartment is an exterior and the floor number. Once we include all variables available, the $R^2$ rises to 0.88 and the Root Mean Square Error (the error between predicted values and actual values) sums to 359 euros. We then test Random Forests and support vector regression (SVR) algorithms, this latter is a common sophisticated machine learning algorithm which uses similar properties to support vector machines for classification predict the values. Preliminary results show the advantage of using Random Forests and SVR for complex models that are more likely to suffer from non-liner relationships between the explanatory variables and the dependent one. \\

The rest of the paper proceeds as follows: the next section offers a description of the related literature. Section 3 describes the data and methods used throughout this work. Section 4 introduces a neighborhood return index. Section 5 shows the rental price estimation using multivariate linear regression and support vector regressions (SVR) accross four different models, and Section 5 offers a preliminary conclusion and steps towards future work on developing the complete index. 

\section{Related literature}

With the ever increasing amount of information about real-sate available online, price prediction of property has become an interesting topic of investigation in recent years. Developments in Machine learning have also enabled such predictions to be made faster and more accurately.\\

In their work, \cite{ma2018estimating} use a dataset of 2,462 warehouse listings in the area of Beijing. Each entry of the dataset contains information about the location, size, distance to city center and second hand house price based on location. With this labelled data, they train four different machine learning algorithms to predict the price of unseen data. From the four models, Linear Regression, Regression Tree, Random Forest Regression and Gradient Boosting Regression Trees,  Random Forest Regression yielded the best performance.
Feature importance describes what information or variables had the greatest impact on predicting the warehouse price. They find that distance from the city centre is the most influential variable, followed by the size of the warehouses and the price of nearby houses.\\

 In order to determine whether the price of real-state display non-linear relationships with the independent variables, \cite{limsombunchai2004house} predict house prices using two models, a Neural Network and a hedonic regression model. They use a set of 200 houses with information such as their size, age, number of rooms, bathrooms, toilets, garages and the availability of amenities on their vicinity. They show that a Neural Network yields more accurate results than the hedonic regression model. This finding is also corroborated in the work of \cite{selim2009determinants}, where they too compare the a hedonic regression model with a Neural Network and find the latter to be more accurate. A reason for this is that there exists heteroscedasticity between house price and the independent variables. This non-linear behavior affects the quality of the predictions and as \cite{ma2018estimating} also show, make non-linear classification models better suited for such tasks. Similar results are also shown by \cite{tabales2013artificial} and \cite{6118994}. \\
 
 Moreover, there is a number of studies that make use of the data from \href{Ideaslita.com}{Ideaslita.com}. For example, \cite{casas2018mirada} download data from this website to analyse the evolution of the real estate market, commercial premises and industrial warehouses market of supply between November 1st, 2016 and May 1st, 2017 in the city of Cordoba, Spain.  Later on, \cite{casas2019mercado} introduce a software for statistical analysis of real estate units built using Java and R for the same city; Cordoba. Their interface displays variables such real estate prices, the geographic location, and several of the characteristics contained in the web-page. \\
 
 Besides, there is a number of studies that focus on real estate prices in a aggregate or macroeconomic set up. For example, \cite{hott2008fundamental} estimate prices based on models of no-arbitrage condition between renting and buying for the USA, UK, Japan, Switzerland and the Netherlands. They find that observed prices deviate substantially and for long periods from their estimated fundamental values in the short run. \cite{born1994real} make use of cycle valuation models that use aggregate cycle measures such as demand and supply cycles, inflation cycles, or rent rate catch-up cycles to evaluate equilibrium real estate prices.

\section{Data and Methods}
 
We use data from Idealista.com which is a real estate platform offering timely data on rents and sales of real estate in Spain, Italy and Portugal.\footnote{\href{https://www.idealista.com/en/data/}{https://www.idealista.com/en/data/}{https://www.idealista.com/en/data/}{https://www.idealista.com/en/data/}} Due to the limited data available to download each month, we narrow down the search to properties in the area of Madrid, and exclude non habitable premises, i.e. offices, garages, premises and warehouses. 

\subsection{Data extraction and dataset creation}
\begin{figure}[h!] \centering
\caption{Transition from JSON file to data-frame}
\includegraphics[scale=0.4]{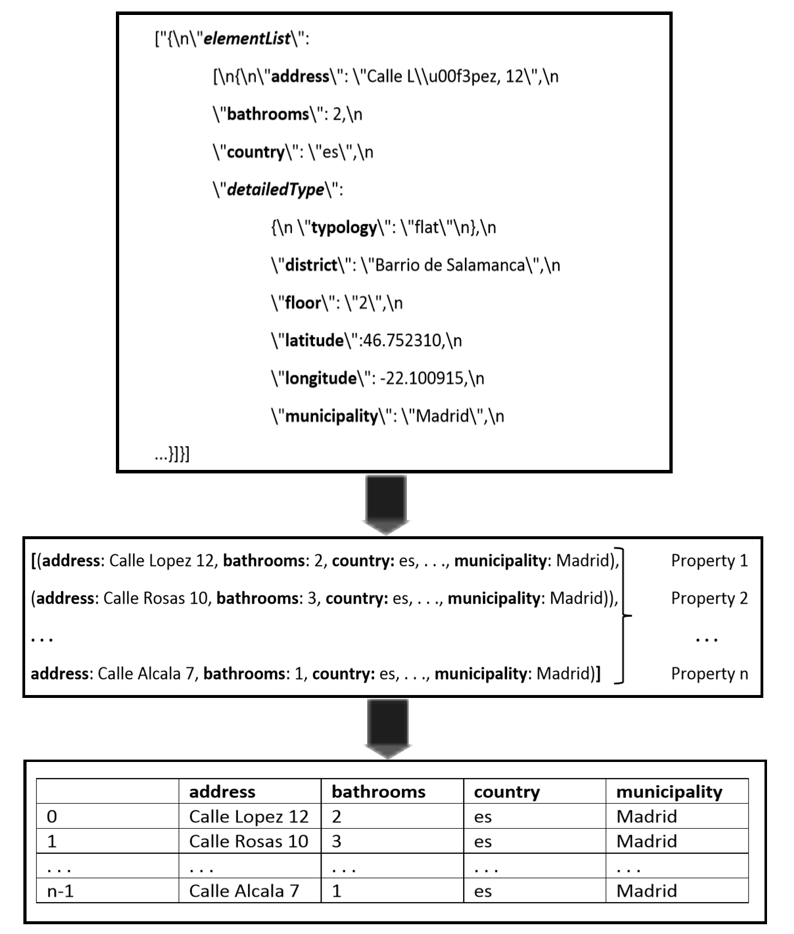}
\label{fig:json_format_example_all}
\begin{flushleft} 
{\small \textbf{Notes:} Example the pipeline used to transform JSON strings into data-frames. On the top, the JSON string returned by the API query. Below, a list of properties extracted using regular expressions operations. On the bottom, the data-frame created with sorted information}	
\end{flushleft}
\end{figure}

In order to download the real state data, we use the API provided by idealista. Idealista provides the user with a password and a key in order to be able to execute queries from their API. An url is created for these queries, specifying several variables that define the characteristics of the search. These variables include operation (rent/sale), center (longitude and latitude coordinates), radius (size of radius of search from center), type of property (house, flat, chalet) etc. Once the query has been executed, the API then returns a JSON file with the retrieved information. An illustration of the road-map can be seen in Figure \ref{fig:json_format_example_all}.\\

In order to be able to work with the extracted data, it must first be prepared and stored in a database. The JSON file downloaded from the API contains the information about all the data points (each individual property) as a long string, where each property, along with all its corresponding information, is enclosed in nested brackets to differentiate it from other data points.\\

Using the \textit{Regex} module of Python to perform regular expressions, the JSON data is cleaned. Brackets, commas, spaces, tabs, new line expressions etc. are first removed, while expressions such as "u00f3" (that in this example represents the accentuated letter \'o) are replaced by their corresponding non-accentuated letters. Once the data is cleaned, each property and its corresponding information is stored as a dictionary. This makes it easier to then use the Python module \textit{Pandas} to create a data-frame from a list of dictionaries containing all data entries. Having a data-frame containing the data facilitates its use for modelling and training purposes.\\

\begin{figure}[h!] \centering
\caption{Descriptive Statistics of sale price}
\includegraphics[scale=0.5]{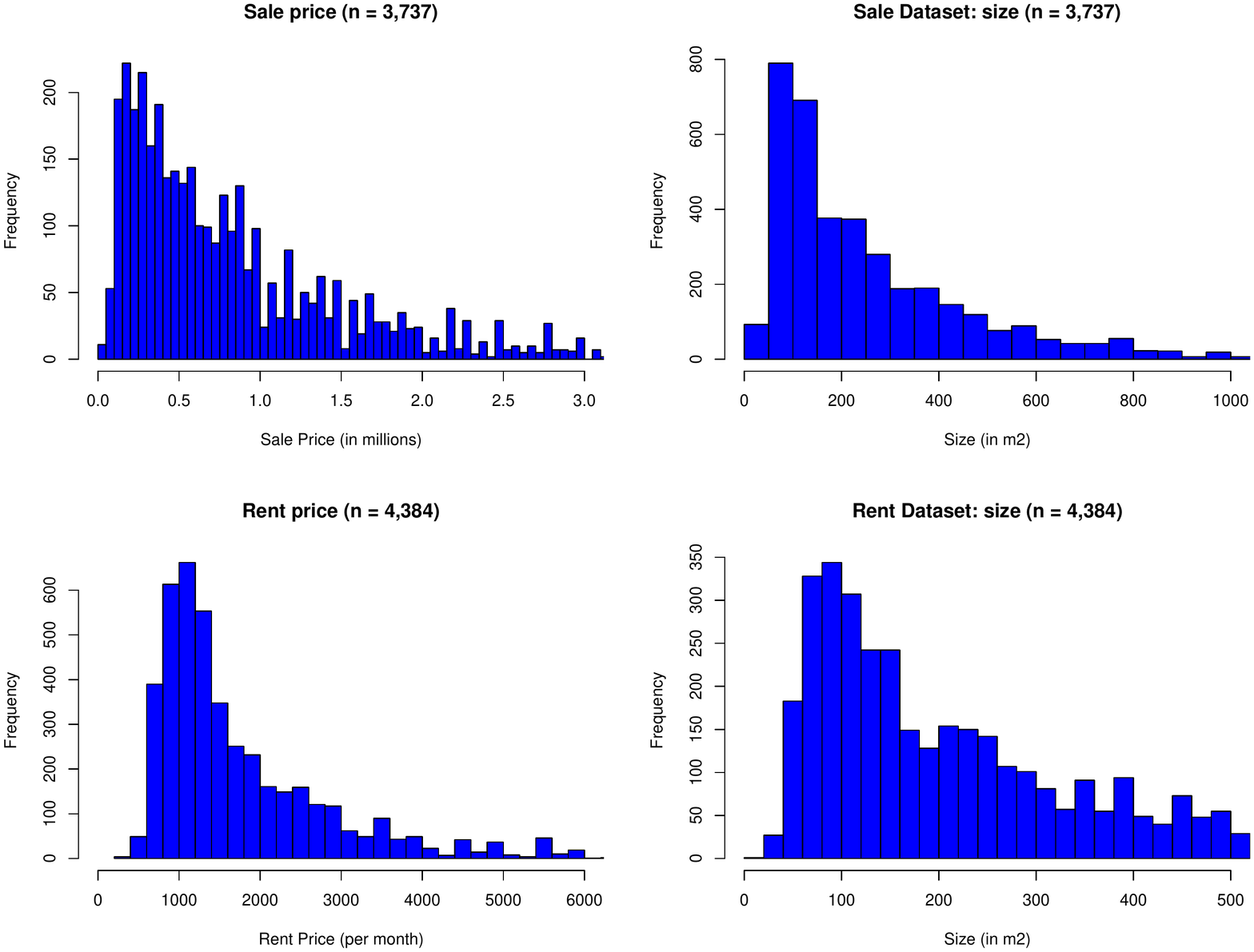}
\label{fig:Descriptive_Statistics}
\begin{flushleft} 
{\small \textbf {Notes:} Top left and bottom left graphs show the distribution of sales and renting prices of houses respectively. The graphs in the top right and bottom right show the size distribution of houses for sale and for rent respectively }	
\end{flushleft}
\end{figure}

To acquire the data trough the API, we set a point along the coordinates and the radius we want to reach. To focus on the Madrid area, we set the central point to 40.4167' and -3.70325" which corresponds to the city centre of the capital and set a radius to 60km. In total, we retrieve 8,121 houses in Madrid where 3,737 are on sale while 4,384 are for rent. Figure \ref{fig:Descriptive_Statistics} shows the distribution in prices for the data extracted. The mean sell price is €970,000 euros while the standard deviation is €1.02 millions. While the average rental price in our sample is 1,912 euros with a 1,683 euros of standard deviation. \\ 

\begin{figure}
\centering
\caption{Example of Geo-located data}
\begin{subfigure}[b]{0.55\textwidth}
   \includegraphics[width=1\linewidth]{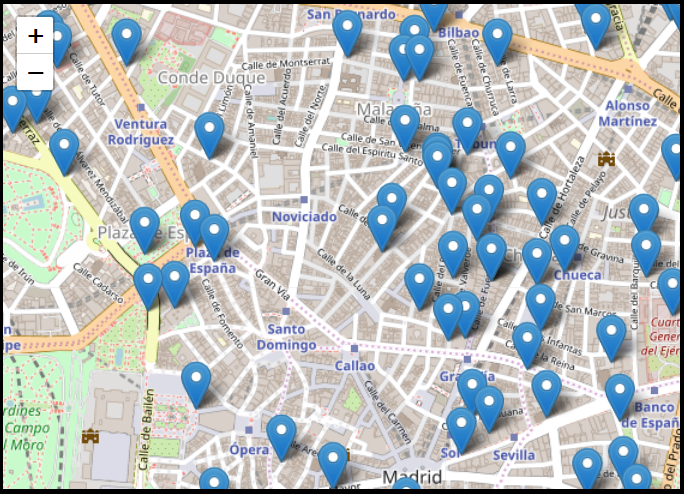}
   \caption{}
   \label{fig:Ng1} 
\end{subfigure}

\begin{subfigure}[b]{0.55\textwidth}
   \includegraphics[width=1\linewidth]{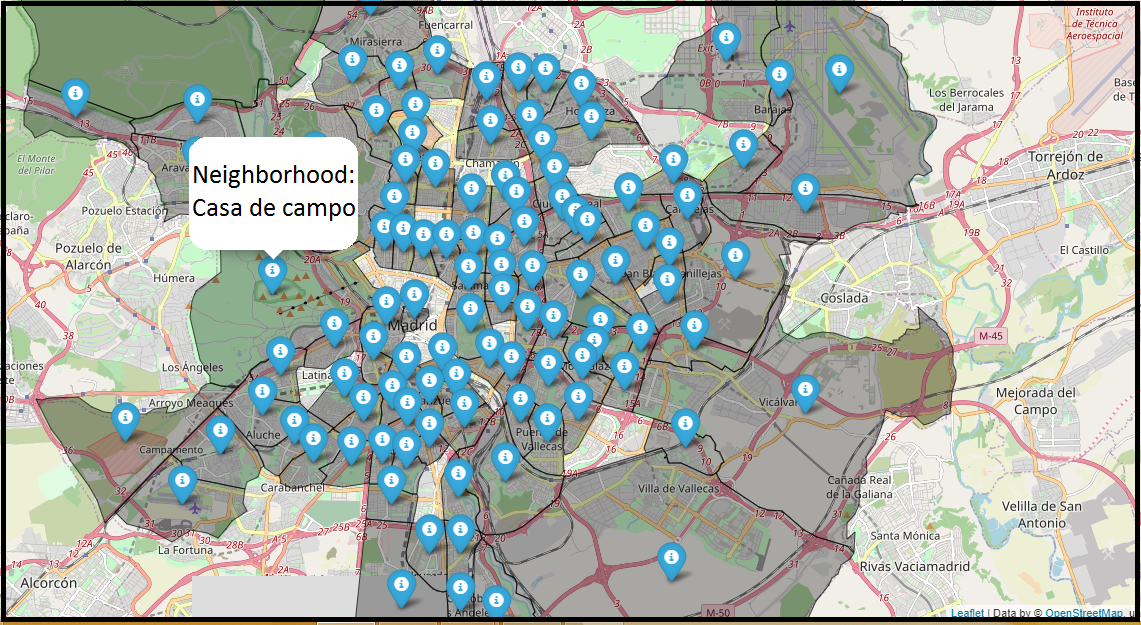}
   \caption{}
   \label{fig:Ng2}
\end{subfigure}

\begin{flushleft} 
{\small \textbf{Notes:} (a) Example of a map in the city centre of Madrid where the blue pop outs are a sample of the location of some houses analysed using the library leaflet (b) Example of a map of Madrid where the blue pop outs are a sample of the location each neighborhood of the Madrid area shaded in dark grey and delimited by dark continuous lines}	
\end{flushleft}
\end{figure}

Figure \ref{fig:Madrid_Map} displays a fraction of houses here analyzed. To build the map we use the longitude and latitude of each house which we can easily locate in the map using the \textit{leaflet} library. Leaflet is one of the most popular open-source JavaScript libraries for interactive maps where one can not only geo-locate objects but which also contains rich information about shops, schools, bus\ stations, gas-stations, and other wide range of services. Using the GEOJSON information about each neighborhood in Madrid, we can extract geo-information about specific areas and also visualize them as shown in figure \ref{fig:Ng2}.\\

\clearpage{}

\section{Neighborhood return}



In this section, the estimation of the sale/rent ratio per neighborhood is presented. This ratio creates an indicator of house profitability, that is, the return of a purchased property in a certain area, given a specific mortgage and assuming the property is rented out. This index is thus calculated using two quantities, the monthly average mortgage per neighborhood, and the monthly average renting price per neighborhood. We take the standard mortgage formula and taking into consideration the transaction costs and down payment we obtain:

\begin{equation}
    M = [(P+c\times P)-0.3\times DP]\frac{r(1+r)^n}{(1+r)^n-1}
\end{equation}

where $M$ is the monthly mortgage sum; $(P+c\times P)$ is the total costs of the property: total price of the house $P$ and the transaction cots $(c\times P)$; $DP$ is the down payment (set to 30\% of the price); $r$ is the monthly interest rates, and $n$ the total number of months that the mortgage will last. In the benchmark set up we set $r$ to $0.0016$, therefore assuming an annual interest rates of 2\% (in line with the monetary policy authority target). Furthermore, we assume a mortgage that will last 30 years, therefore $n = 360$; and the transaction costs $c$ to be 6.7\% of the total price. Following this formula we would obtain a monthly payment during 30 years of 423 euros for a unit of real estate that costs 150,800 euros (or 160,903 euros once the transaction costs are taken into account). \\

Our data contains houses that range from 30${m}^2$ to over 1,200${m}^2$, and there is some disparity in the distribution of the houses for sale and for rent with respect to size, at it can be seen in figure \ref{fig:Descriptive_Statistics}. In the graph of the size distribution of the houses for sale, it can be seen that the distribution does not drop as abruptly as the size increases, as it does for the rental distribution. A reasonable assumption for this disparity is that it is more likely that a house in a wealthy suburb of Madrid with 1,200${m}^2$ is sold than rented out. Furthermore, house sale and rental prices can show heteroskedasticity with respect to size. Thus, the standard deviation across the entire size spectrum will be larger for houses for sale than for rent.\\

In order to account for these differences in the size distribution of property for sale and rent we calculate the average mortgage payed per neighborhood and for a given size interval. For example, an instance of this calculation would be the average mortgage payed in the Neighborhood of "Prosperidad" for all houses that have a size between 30${m}^2$ and 60${m}^2$, which is calculated using 36 samples and yields 1,581.86€ per month. Then, the average rent per neighborhood and for the different size intervals is also calculated. In the previous example, the average monthly rent payed in "Prosperidad" for properties between 30${m}^2$ and 60${m}^2$ is of 1,371.95€ per month, calculated with 41 samples. By dividing the average monthly rent by the average monthly mortgage of the instances that belong to the same neighborhood and size interval, an index of the profitability or return of real state in a given area and for a given size range is obtained. Using again the example of "Prosperidad" and properties of a size between 30${m}^2$ and 60${m}^2$, this index is 0.867. Generally, a value of 1 would be expected, since the renting prices are adjusted to the housing prices over time. Thus,  the higher the index value, the higher the return of real state since there is a larger gap between the rental prices minus the mortgage prices. An index value smaller than 1 like the one of the example points at an area with negative return. \\

\begin{figure}[h!] \centering
\caption{Sample of proportion of houses` geo-location}
\includegraphics[scale=0.20]{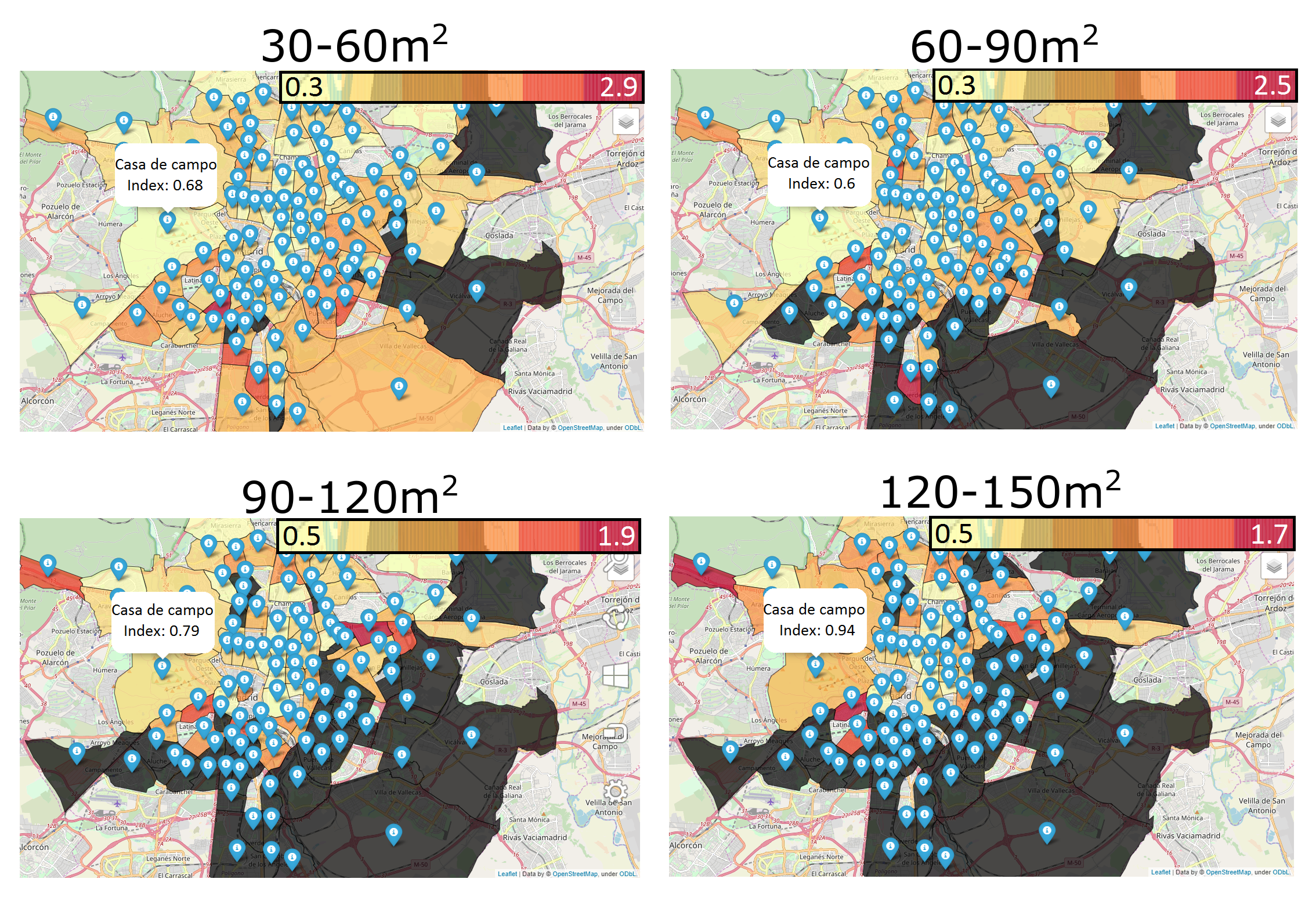}
\label{fig:Madrid_Map}
\begin{flushleft} 
{\small \textbf{Notes:} Profitability index shown for each neighborhood for different size intervals of real state }	
\end{flushleft}
\end{figure}

Table \ref{tab:ReturnIndex_Neiborhood} shows this index computed for each neighborhood and size, and Figure \ref{fig:Madrid_Map} shows the same index as a heat map for all neighborhoods of Madrid with available data and for four different size ranges. As it can be seen from Figure \ref{fig:Madrid_Map}, there are more houses in the market with a size range of 30${m}^2$ and 60${m}^2$. The supply of houses decreases for bigger size ranges, which can be seen in the grayed out neighborhoods of in the figure, where no data points were available.\\

\clearpage
\section{Rent price prediction models}

In this section, we will like to explore a different approach into obtaining returns for each neighborhood. So far we have been looking at the differences between the average price of real estate for sale (and therefore their average monthly mortgage rate) and the average rental price of real estate for rent per neighborhood. For this approach to be valid, we have to assume identical characteristics between the rental and sale real estate markets per neighborhood (neighborhood homogeneity). \\

With this in mind, we now want to train several classification models which will assess what is the renting price that a real estate for sale is likely to obtain. The classifications models will be trained in the set of rental property in order to learn which specific characteristics explain their monthly price. We will then evaluate the performance of each model (accuracy) by splitting the sample into training and testing data1 sets. After this evaluation we will take each of the real state units for sale, and calculate their rental price using the model that yielded the best results. We can therefore obtain the rental market price that a real estate unit for sale will get given its characteristics. Nonetheless, we will have to assume that hidden characteristics of the rental set (e.g. age) is similar to that of the selling set, although this no longer has to hold at the neighborhood level.\\

We use the rental price per month as our variable to predict based on several characteristics of the real estate unit. These characteristics are: I) the average price by area, II) the floor in which the real estate unit is located (empty for houses), II) whether or not is an exterior or interior flat (facing the main street or, on the contrary, an inner patio), III) whether or not the building has a lift, IV) whether or not it includes parking, V) whether or not is a new development, VI) The number of photos (proxy for the interest into selling the flat), VII) property type (chalet, duplex, flat or penthouse), VIII) the size in square meters, IX) the status (good, new development, or renewed), and X) the number of bathrooms divided by the number of square meters.\footnote{We divide the number of bathrooms per the size of the real estate unit in order to remove multicollinearity among the two variables: the bigger the flat, the more likely it is to have more bathrooms.} As our benchmark model we will use a simple linear regression to predict house prices. We will then use Random forest which have become a very popular “out-of-the-box” learning algorithm that enjoys good predictive performance. And finally, we will use a Support Vector Regressions (SVR).

\subsection{Multivariate linear regression}

We start our analysis with the simplest of all forms: multivariate linear regression. Our dependent variable is the rental price per square meter and the controls or independent variables are the variables previously mentioned. Recall that in linear regression, the relationships are modeled using linear predictor functions whose unknown model parameters are estimated from the data. Linear prediction functions is the best fitted line using Ordinary Least Squares (OLS) criterion which minimizes the sum of squared prediction error. In other words, we are imposing linear associations between the control variables and our dependent variable; rental price. \\

Table \ref{tab:multilinearregression} displays the results of several multi-variable regressions that we performed on rental price. We perform several regressions in order to check for possible multicollinearity among our regressors. We start with a simple regression consisting only on the size (in $m^2$), whether or not the apartment is an exterior (versus interior) and the floor (column 1).  The adjusted R-squared of this simple regression indicates a good fit: 0.62 indicating that 62\% of the variability in the dependent variable is explained by these controls alone. These results indicate that for every additional square meter, the rental price increases 11 euros on average. In addition, exterior flats are on average 176 euros more expensive to rent (\textit{ceteris paribus}), and for every additional floor, rental price increases 15 euros on average. In Column 2 we introduce whether or not the flat has a lift, indicating that if that is the case, the rental price on average is 98 euros more expensive. In the next column, we introduce price per area, a continuous variable that captures the average square meter rental price per neighborhood. It takes an average of 16 and a maximum value of 84. This variable is very significant and its coefficient states that for every additional euro per square meter on price by area, the rental price increases by almost 90 euros. This highlights the heterogeneity of rental prices per area in Madrid. \\ 

\begin{table}[!htbp] \centering \small
\renewcommand{\arraystretch}{0.6}
  \caption{Linear regression Results} 
  \label{tab:multilinearregression} 
\begin{tabular}{@{\extracolsep{-7pt}}lD{.}{.}{-3} D{.}{.}{-3} D{.}{.}{-3} D{.}{.}{-3} } 
\\[-1.8ex]\hline 
\hline \\[-1.8ex] 
 & \multicolumn{4}{c}{\textit{Dependent variable:}} \\ 
\cline{2-5} 
\\[-1.8ex] & \multicolumn{4}{c}{Rental price} \\ 
\\[-1.8ex] & \multicolumn{1}{c}{(1)} & \multicolumn{1}{c}{(2)} & \multicolumn{1}{c}{(3)} & \multicolumn{1}{c}{(4)}\\ 
\hline \\[-1.8ex] 
Intercept & 145.455^{***} & 256.191^{***} & -1,626.573^{***} & -2,706.799^{***} \\ 
  & (56.163) & (53.753) & (39.964) & (147.025) \\ 
  & & & & \\ 
Size & 10.869^{***} & 13.380^{***} & 15.273^{***} & 14.377^{***} \\ 
  & (0.157) & (0.216) & (0.127) & (0.227) \\ 
  & & & & \\ 
Exterior & 176.281^{***} & -191.876^{***} & 112.010^{***} & 72.632 \\ 
  & (46.443) & (38.112) & (22.294) & (70.331) \\ 
  & & & & \\ 
Floor & 14.493^{***} & 7.074^{***} & 0.987 & 0.196 \\ 
  & (2.228) & (1.641) & (0.947) & (1.790) \\ 
  & & & & \\ 
Lift &  & 98.001^{**} & 84.459^{***} & 376.903^{***} \\ 
  &  & (42.355) & (24.357) & (84.557) \\ 
  & & & & \\ 
Price by Area &  &  & 89.578^{***} & 130.930^{***} \\ 
  &  &  & (1.205) & (3.079) \\ 
  & & & & \\ 
Status."good" &  &  &  & 56.697 \\ 
  &  &  &  & (90.208) \\ 
  & & & & \\ 
Status."New Development" &  &  &  & 48.821 \\ 
  &  &  &  & (440.670) \\ 
  & & & & \\  
Status."renew" &  &  &  & 207.371 \\ 
  &  &  &  & (264.721) \\ 
  & & & & \\ 
 Bathrooms\_sqm &  &  &  & 4,121.352 \\ 
  &  &  &  & (3,610.322) \\ 
  & & & & \\ 
 duplex &  &  &  & 1.343 \\ 
  &  &  &  & (64.636) \\ 
  & & & & \\ 
 flat &  &  &  & 66.751 \\ 
  &  &  &  & (48.698) \\ 
  & & & & \\ 
Parking &  &  &  & 115.252^{***} \\ 
  &  &  &  & (39.512) \\ 
  & & & & \\ 
Photos &  &  &  & 2.750^{**} \\ 
  &  &  &  & (1.170) \\ 
  & & & & \\ 
\hline \\[-1.8ex] 
Observations & \multicolumn{1}{c}{3,026} & \multicolumn{1}{c}{2,734} & \multicolumn{1}{c}{2,734} & \multicolumn{1}{c}{946} \\ 
R$^{2}$ & \multicolumn{1}{c}{0.624} & \multicolumn{1}{c}{0.606} & \multicolumn{1}{c}{0.870} & \multicolumn{1}{c}{0.886} \\ 
Adjusted R$^{2}$ & \multicolumn{1}{c}{0.624} & \multicolumn{1}{c}{0.605} & \multicolumn{1}{c}{0.869} & \multicolumn{1}{c}{0.884} \\ 
\hline 
\hline \\[-1.8ex]
\end{tabular} 
  \begin{flushleft} 
{\small\textit{Note:}  In this table, we regress rental price on several real state characteristics variables. Status indicates the description that the owner gives to the property; good, new development, or renewed. Standard errors are reported in parentheses. *, **, and *** indicate statistical significance at the 10\%, 5\%, and 1\% level, respectively.} \\ 
  \end{flushleft}
\end{table}

Finally, column 4, includes several other reggressors such as the status (as a dummy variable), number of bathrooms per square meter, the number of photos attached to the ad, and whether or not parking is included in the price. Here we see that for each additional photo, the price rises by two euros. Given that we are controlling for size (we would expect a larger amount of photos for bigger units), this seems to indicate a that the more photos placed by the owner, the higher the asking price. In addition, and as expected, those units with parking are on average 115 euros more expensive. It is interesting to see that as we introduce more controls variables, such as in columns 3 and 4, the intercept becomes negative. This might be a sign of non-linearities in our data set. For this reason, it might be important to incorporate additional non-linear statistical tools to study the dynamics behind rental prices. \\

\begin{figure}[h!] \centering
\caption{Multivariate linear regression predictors}
\includegraphics[scale=0.55]{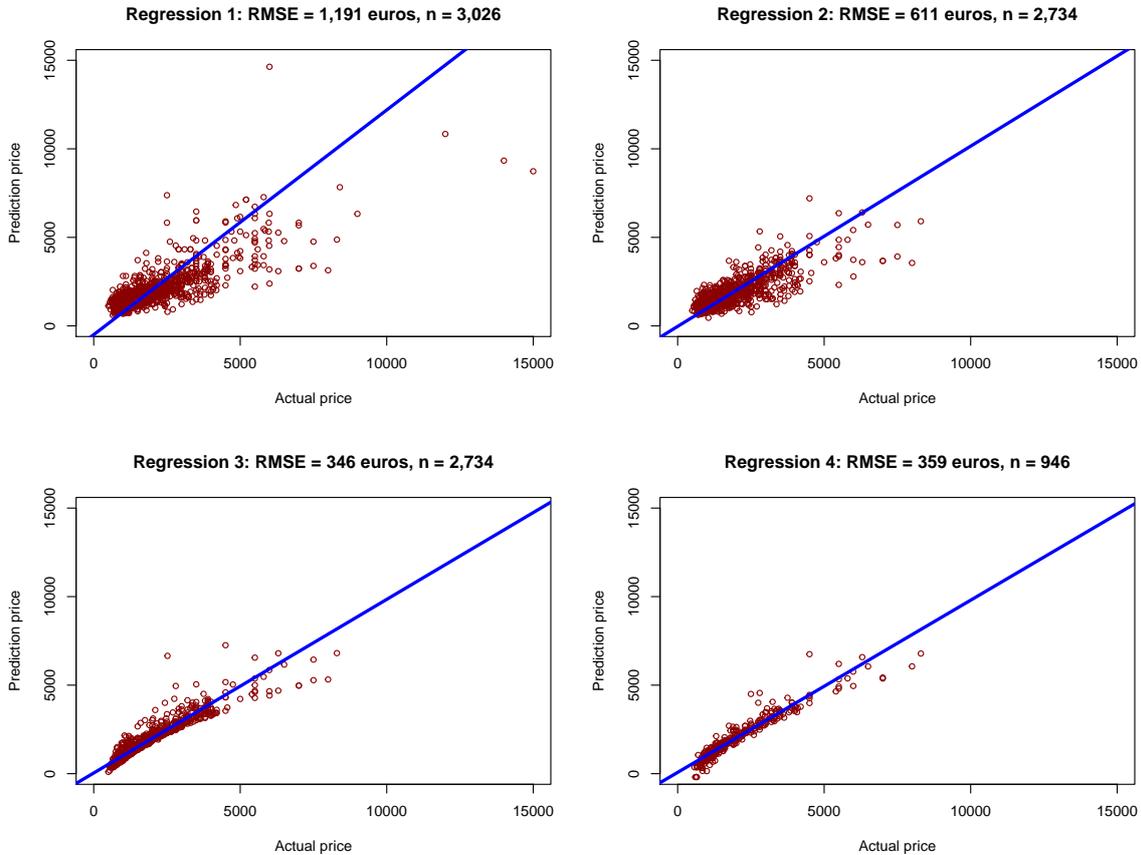}
\label{fig:LinearRegression_Prediction}
\begin{flushleft} 
{\small \textbf{Notes:} scatter plots between the predicted price (y-axes) and the actual price (x-axes) for each of the four regressions in Table \ref{tab:multilinearregression}}	
\end{flushleft}
\end{figure}

Figure \ref{fig:LinearRegression_Prediction} displays the prediction price over the actual price for each of the four multivariate linear regressions. To run the prediction we split the sample randomly into training (70\% of the data) and testing (30\% of the data). As we can see, there is a much better fit (lower square mean squared error) for regression 4 than for regression 1; 359 v.s 1,191. Nonetheless, there are a few drawbacks that prevent us from concluding that regression 4 is superior to regression 1. On the one hand, we do not have the same number of observations for the two regressions, i.e. only 946 observations for regression 4 whereas regression 1 is built from 3,026 observations. Moreover, as stated before, as we include more observations, non-linearities between the real estate characteristics and rental price become stronger. To illustrate this point, compare the results from regression 2 and regression 3: even through they have the same number of observations, the prediction on average is lower for the model with fewer controls. Nonetheless, whereas the error in prediction is more or less constant for the whole price range in regression 2, that for regression 3 is very low for cheaper flats while increases as price becomes more expensive. In other words, as the price increases we keep under-predicting its price. 

\subsection{Random Forests}

Random forests are built on the same fundamental principles as decision trees and bagging. The concept of bagging, also known as bootstrap aggregation, is to create several subsets of the data by randomly selecting data-points from the training data with replacement (some observations may be repeated). Decision trees are then trained with the subsets, and their results averaged. Bagging trees introduce a random component into the tree building process that reduces the variance of a single tree’s prediction and improves predictive performance. However, the trees in bagging are not completely independent of each other since all the original predictors are considered at every split of every tree. Rather, trees from different bootstrap samples typically have similar structure to each other (especially at the top of the tree) due to underlying relationships. \\

For example, if we create several decision trees with different bootstrapped samples in our set, most likely all trees will have a very similar structure at the top.\footnote{see for example \href{https://uc-r.github.io/random_forests}{https://uc-r.github.io/random$\_$forests}} This characteristic is known as tree correlation and prevents bagging from optimally reducing the variance of the predictive values. In order to further reduce the variance, we need to minimize the amount of correlation between the trees. This can be achieved by injecting more randomness into the tree-growing process. Random forests achieve this in two ways. On the one hand, we can use the \textit{Bootsrap re-sampling process}, where each tree is grown to a bootstrap re-sampled data set in order to de-correlated them. On the other hand, we can use the \textit{Split-variable randomization process} where the search in the split variable is limited to a random subset of $m$ of the $p$ variables.\footnote{For regression trees, typical default values are $m = \frac{p}{3}$ but this should be considered a tuning parameter.} \\



Random forests have a handfull of hyperparameters that need to be tuned during training. Typically, the primary concern when starting out is tuning the number of candidate variables to select from at each split and few additional parameters: the number of trees; the number of variables to randomly sample as candidates at each split; the number of samples to train on; the minimum number of samples within the terminal nodes; and the maximum number of terminal nodes. \\

For our training, we focused on tuning the number of candidate variables and the number of trees, setting the rest of the parameters to constant values. In the case of the trees depth, we set the model to expand them until all the leaves are pure. The number of variables to randomly sample at each split is set randomly with a given seed so that the results are always reproducible. When training the model, it was observed that outliers in the data (e.g. houses with rental prices of 30,000€ or houses with a size of 10,000${m}^2$) greatly impacted the results. For this reason, we implemented outlier elimination using Z-Scores.\\

\begin{figure}[h!] \centering
\caption{Random Forest predictors}
\includegraphics[scale=0.55]{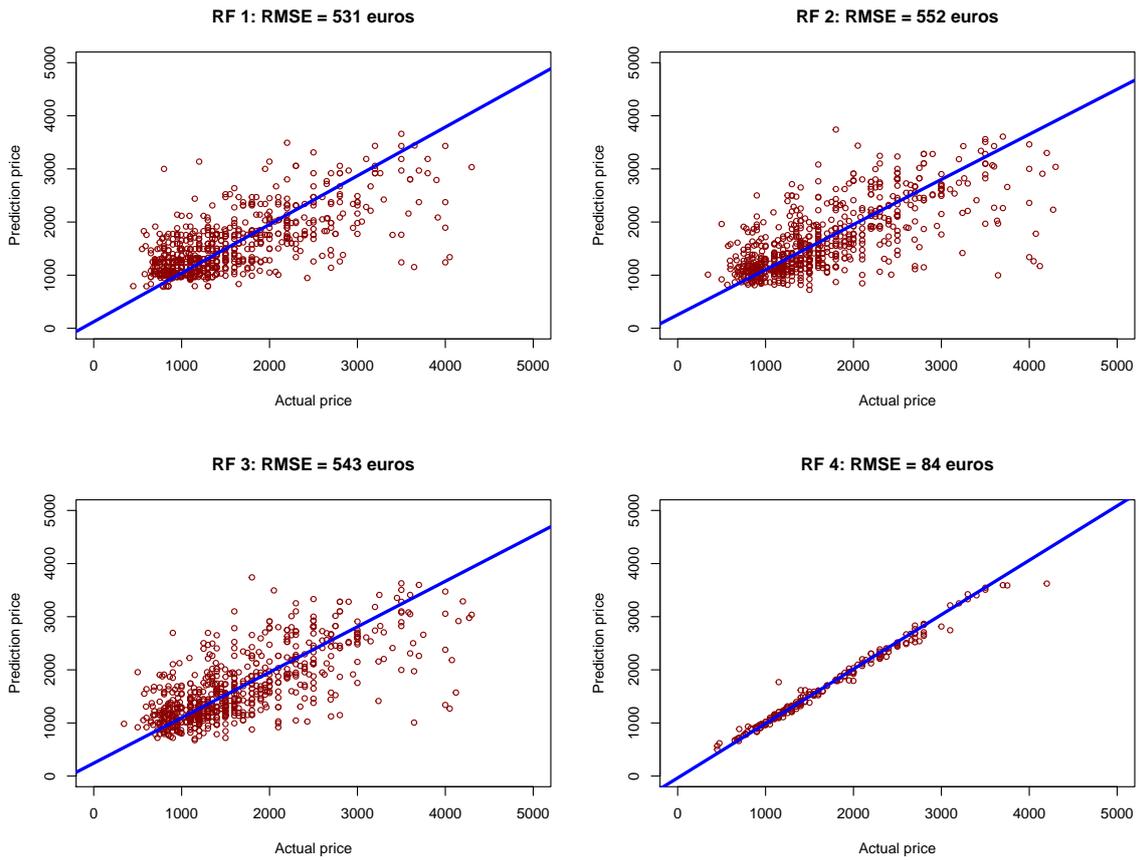}
\label{fig:RF_Prediction}
\begin{flushleft} 
{\small \textbf{Notes:} scatter plots between the predicted price (y-axes) and the actual price (x-axes) for each of the four regressions in Table \ref{tab:multilinearregression}. Training done with 100 trees. From top left to lower right, ${1}^{st}$ model trained with 3652  samples, $2^{nd}$ and $3^{rd}$ with 3598	and $4^{th}$ with 1212}
\end{flushleft}
\end{figure}

In order to find the best combination of number of trees, number of candidate variables and size of the Z-score, we performed a grid search spanning 10 to 500 trees, four to ten variables and 0.5 to 10 Z-score. Results consistently showed that a performance peak was achieved when using between 100 to 125 decision trees, and a Z-score of 1.5. Figure \ref{fig:RF_Prediction} shows the results of the tests of the the model on four different variable subsets, trained with 100 decision trees and a Z-score of 1.5. The subsets of variables used for the training are the same as those used for the training of the multi-linear regressor shown in Table \ref{tab:multilinearregression} of the previous section. The first model was thus trained with size (in ${m}^2$), whether the apartment is exterior or not, and what floor it is at. For the second training we introduce whether or not the house has a lift. For the third, price by area is taken into account, and the the last training, the status of the house (good, new development or renewed), whether or not it has a parking space, the type of house (duplex, flat), and the number of bathrooms per squared meter. Due to incomplete data, the more variables are introduced, the less complete data samples are. Thus, models that include more training variables are trained with less data. Figure \ref{fig:RF_Prediction} shows how the results dramatically improve on the fourth training, where more explanatory variables are introduced, yielding an impressive RMSE of 84€. This points to the fact that the state of the house, as well as its type and the number of bathrooms per square meter play a crucial role in determining their rental price. The fourth model is trained with 1,212 samples, as opposed to the first three models that are trained with over 3,500 samples each. However, there is no indication that this smaller subset is skewed or falls outside the larger super-sets, since all the common variables have roughly the same distribution.


\clearpage{}
\subsection{Support Vector Regressions}

\begin{figure}[h!] \centering
\caption{Support Vector Machines}
\includegraphics[scale=0.3]{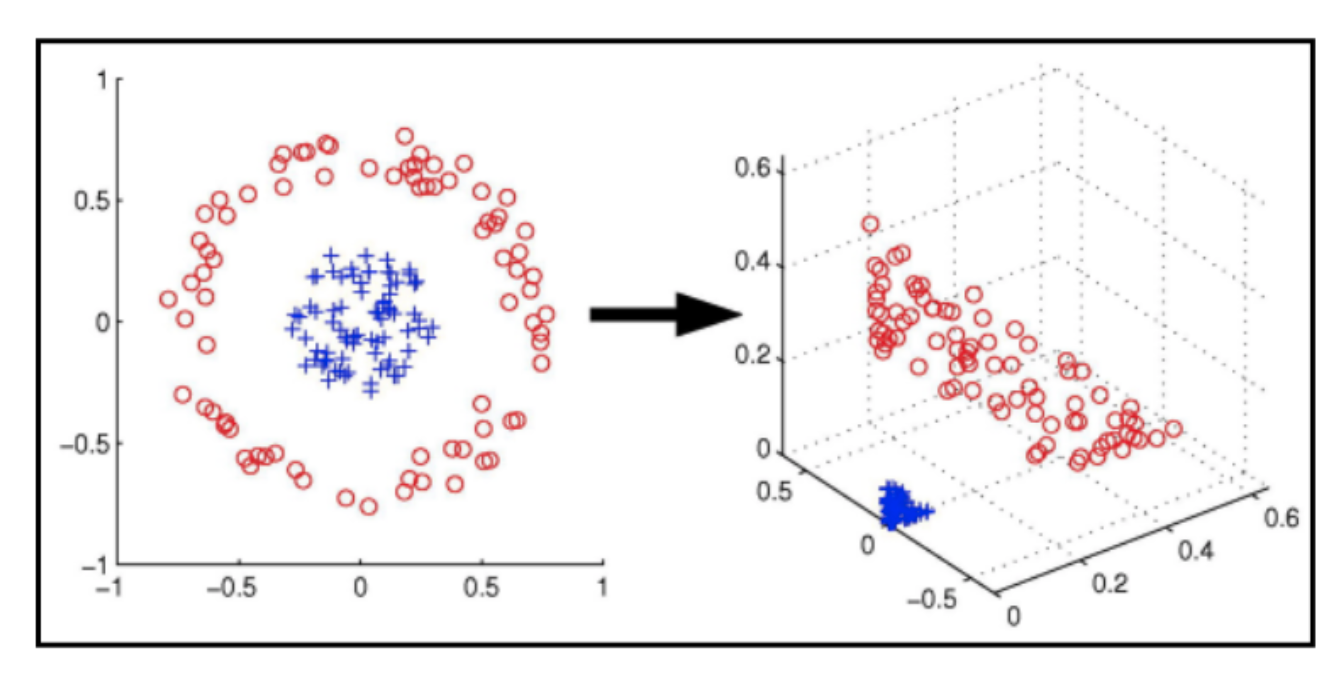}
\label{fig:SVM_Illustration}
\begin{flushleft} 
{\small \textbf{Notes:} Image obtained from \href{https://medium.com/@zachary.bedell/support-vector-machines-explained-73f4ec363f13}{https://medium.com/@zachary.bedell/support-vector-machines-explained-73f4ec363f13}}	
\end{flushleft}
\end{figure}

Support vector machine (SVM) analysis is a popular machine learning tool for classification and regression, developed in 1992 by \cite{boser1992training}. SVM belong to the family of generalized linear classifiers in the sense that it is a prediction tool that uses machine learning theory to maximize predictive accuracy while automatically avoiding over-fitting to the data (\cite{jakkula2006tutorial}). Support vector machines can be defined as systems which use hypothesis space of a linear functions in a high dimensional feature space, trained with a learning algorithm from optimization theory that implements a learning bias derived from statistical learning theory. To illustrate this point, consider the data point presented in Figure \ref{fig:SVM_Illustration} where we have an inner circle of data that belongs to a cluster while an outer data points that belong to a different cluster. It will be impossible to correctly classify both clusters by tracing a line. For this reason, we incur to a 3D transformation of the data points (right panel), where we can with no major complications depict the two data clusters using only a straight line (or more specifically, a linear plane).  \\

Support Vector Regression (SVR) works on similar principles as Support Vector Machine (SVM) classification in the sense that SVR is the adapted form of SVM when the dependent variable is numerical rather than categorical. The main benefit of using SVR (or SVM) is that it is a non-parametric technique, therefore we do not assume certain conditions or parameters in the data (e.g. linear combinations or lack of heteroskedasticity in our sample). SVR uses the principle of maximal margin, meaning that we do not care so much about the prediction as long as the error term ($\epsilon$) is less than a certain value. In other words, maximal margin allows viewing SVR as a convex optimization problem. Besides, the regression can also be penalized using a cost parameter (later explain in more detail), which helps to avoid over-fitting. \\

Given that we are not using too many features (control variables) to predict the rental price, we do not incur in feature reduction methods such as Recursive Feature Elimination (RFE) or Principal Components Analysis (PCA). Neither we use in log price-transformation, something which might increase the accuracy of the model. In this sense, we feed raw data to the support vector regressions and test for the accuracy using the Root Mean Square Error (RMSE). Just as before, we split the data into training and testing (0.7 and 0.3 respectively) and apply the classification test on the testing data set only. We consider four different kernel functions to map a lower dimensional data into a higher dimensional one: i) linear: $u'v$, ii) polynomial: $(\gamma u'v + coef0)^{degree}$, iii) radial basis: $e^{(-\gamma | u - v |^2)}$ and sigmoid: $tanh(\gamma u'v + coef0)$, where $u'$ and $v$ are the vectors representing the inputs in the vector space and $\gamma$ is a weighting factor that scales the amount of influence that two data points have on each other. Besides, we set the cost of constraints violation to the default $cost=1$. This is the 'C'-constant of the regularization term in the Lagrange formulation, or put it in other words, by how much you want to avoid misclassifying in each training example. For large values of C, the optimization will choose a smaller-margin hyperplane of that hyperplane that does a better job of getting all the training points classified correctly. Finally, note that the number of support vectors are given by the model and usually ranges from 260 to 2,900. \\

\begin{figure}[h!] \centering
\caption{Support Vector Regressions RMSE}
\includegraphics[scale=0.55]{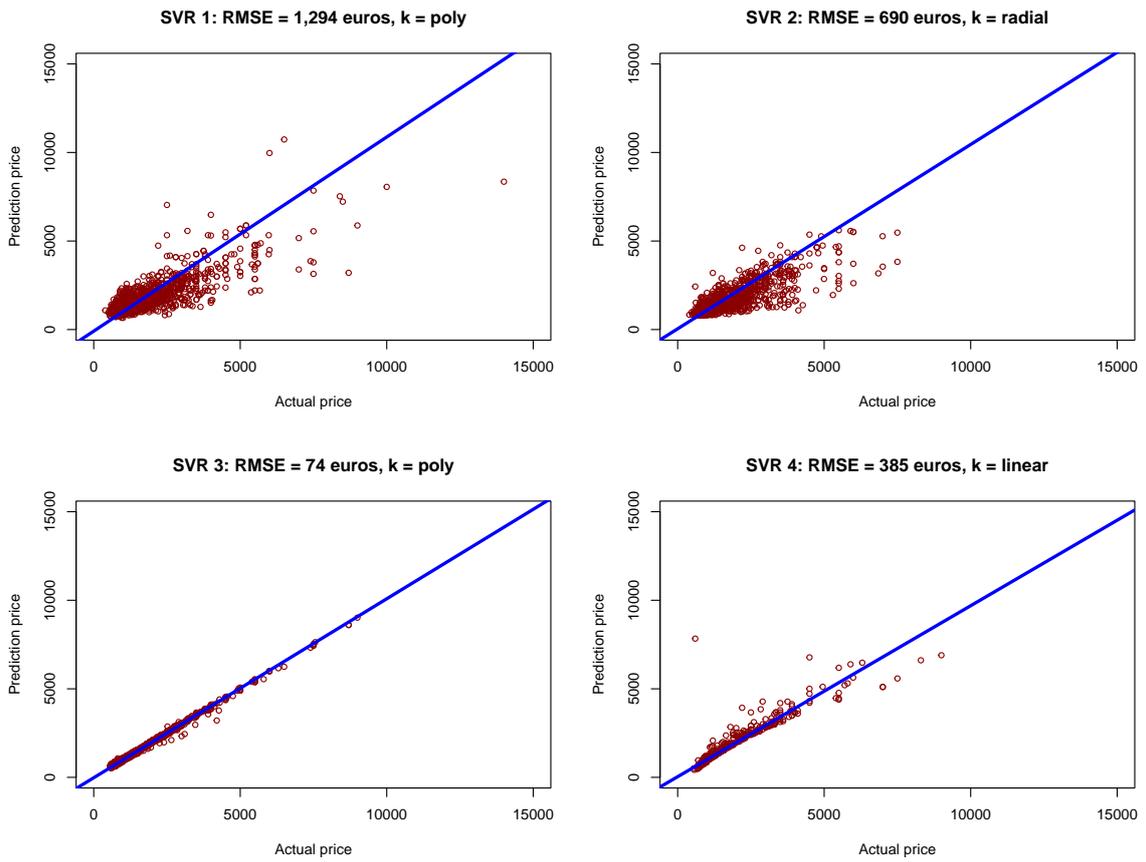}
\label{fig:SVR_Results_Figure}
\begin{flushleft} 
{\small \textbf{Notes:} scatter plots between the predicted price (y-axes) and the actual price (x-axes) for each of the four support vector regressions considered, see \ref{tab:SVR_Results}. K stands for the type of kernel used (better prediction power). }	
\end{flushleft}
\end{figure}

Table \ref{tab:SVR_Results} shows the RMSE across each of the four models and four different kernel functions. For the first model, where we only have the $size$, $exterior$ and $floor$ as dependent variables, the RMSE using a polynomial kernel is 1,191. This is the same error than when using a multivariate-linear regression approach. Once we include the variable $lift$ in our model, the RMSE drops to 690 under the radial basis kernel which is slightly less accurate than the multivariate linear regression: 611. The next specification adds price by area and is in this model when we start seeing a much higher accuracy for SVR compared to the multivariate linear regression. Under the polynomial kernel, the RMSE turns out to be only 74 whereas that for the multivariate linear regression was 346. Recall that the multivariate linear regression kept underpredicting the price for the most expensive units which illustrate the limitations to adjust non-linearities in the data. This is not the case of the SVR, which indepndently of the price, the estimation runs through a straight line (see Figure \ref{fig:SVR_Results_Figure}). This is also the case for the last specification, where we incorporate all variables that we have. Although the RMSE is slightly higher for the SVR than the multivariate linear regression; 385 and 359 respectively, the prediction price and the actual price tend to lie in a straight line (see bottom-right panel of Figure \ref{fig:SVR_Results_Figure}). All in all, we can conclude that the multivariate linear regression performs better for simpler models with fewer variables, as it will not be able to capture non-linearities in the data. \\

\begin{table}[htbp]
  \centering
  \caption{Support Vector Regression Results}
    \begin{tabular}{lrrrr}
          & \multicolumn{4}{c}{\textbf{Root Mean Square Errors (RMSE) }} \\
\cmidrule{2-5}    \textbf{Kernel} & \multicolumn{1}{l}{\textit{\textbf{SVR 1}}} & \multicolumn{1}{l}{\textit{\textbf{SVR 2}}} & \multicolumn{1}{l}{\textit{\textbf{SVR 3}}} & \multicolumn{1}{l}{\textit{\textbf{SVR 4}}} \\
    \textit{\textbf{linear}} & 1,255 & 706   & 392   & 385 \\
    \textit{\textbf{polynomial}} & 1,194 & 702   & 74    & 13,401 \\
    \textit{\textbf{Radial}} & 1,294 & 690   & 154   & 1,216 \\
    \textit{\textbf{Sigmoid}} & 145,155 & 44,193 & 55,453 & 1,218 \\
    \end{tabular}%
  \label{tab:SVR_Results}%
\begin{flushleft} 
{\small \textbf{Notes:} This table displays the Root Mean Square Errors of the different models and kernels using Support Vector Regression. Note that the additional model parameters such as cost, gamma, epsilon are default values. }	
\end{flushleft}
\end{table}%

\clearpage{}

\section{Conclusions and future work}

In this preliminary work, we have motivated a tool for investment in real estate units. In particular, we pave the road towards developing a rental yield algorithm to infer the most likely rental price of each real estate unit for sale. The rental yield will be the difference between the monthly mortgage payment of the unit and the most likely monthly rental price. In this project we present a preliminary work on how we gather the data and the type of algorithms that we use. For future work, we would like to expand the number of control variables using geo-location data such as closeness to sport, shopping areas, transport or medical facilities.  

\clearpage


\bibliography{bibliografia} 
\bibliographystyle{apalike}


\section{APPENDIX}

\begin{table}[] \small \renewcommand{\arraystretch}{0.4}
\caption{Return Index per neighborhood and size}
\label{tab:ReturnIndex_Neiborhood}
\begin{tabular}{llllllll}
                                         & \multicolumn{5}{c}{Return Index per size}                      &  &         \\
Neighborhood                             & (30-60) & (60-90) & (90-120) & (120-150) & (\textgreater{}150) &  & Average \\
12 de Octubre-Orcasur                    & 2.33    & 2.33    &          &           &                     &  & 2.33    \\
Abrantes                                 & 1.63    & 1.46    &          &           &                     &  & 1.55    \\
Acacias                                  & 1.03    & 1.08    & 1.56     &           &                     &  & 1.22    \\
Adelfas                                  & 1.06    & 1.13    & 1.02     &           &                     &  & 1.07    \\
Aguilas                                  & 1.77    &         &          &           &                     &  & 1.77    \\
Alameda de Osuna                         & 0.85    & 0.66    & 0.58     &           &                     &  & 0.70    \\
Almagro                                  & 0.57    & 0.59    & 0.58     & 0.61      & 0.62                &  & 0.59    \\
Aluche                                   & 1.80    &         &          &           &                     &  & 1.80    \\
Apostol Santiago                         & 1.03    & 0.79    & 0.79     &           &                     &  & 0.87    \\
Arapiles                                 & 0.67    & 0.71    & 0.76     & 0.72      & 0.72                &  & 0.72    \\
Aravaca                                  & 0.90    & 0.92    & 0.96     & 1.09      & 1.12                &  & 1.00    \\
Arroyo del Fresno                        & 0.88    & 0.88    & 0.88     & 0.74      & 0.74                &  & 0.83    \\
Atalaya                                  & 0.34    & 0.34    &          &           &                     &  & 0.34    \\
Bellas Vistas                            & 1.92    & 1.85    &          &           &                     &  & 1.88    \\
Bernabeu-Hispanoamerica                  & 0.70    & 0.78    & 0.88     & 0.88      & 0.89                &  & 0.83    \\
Berruguete                               & 1.53    & 1.29    &          &           &                     &  & 1.41    \\
Buena Vista                              & 2.00    &         &          &           &                     &  & 2.00    \\
Butarque                                 & 1.45    &         &          &           &                     &  & 1.45    \\
Campamento                               & 0.71    & 0.79    &          &           &                     &  & 0.75    \\
Campodelas Naciones-Corralejos           & 0.97    & 0.91    & 0.92     & 0.92      & 0.92                &  & 0.93    \\
Canillas                                 & 0.66    & 0.71    & 0.82     & 1.11      & 1.11                &  & 0.88    \\
Canillejas                               & 0.80    & 0.85    & 1.49     & 1.07      & 1.07                &  & 1.05    \\
Casa de Campo                            & 0.68    & 0.60    & 0.79     & 0.94      & 0.94                &  & 0.79    \\
Casco Antiguo                            & 1.45    & 0.90    & 0.87     &           &                     &  & 1.07    \\
Casco Historico de Barajas               & 1.31    &         &          &           &                     &  & 1.31    \\
Casco Historico de Vallecas              & 1.51    &         &          &           &                     &  & 1.51    \\
Castellana                               & 0.57    & 0.60    & 0.64     & 0.72      & 0.76                &  & 0.66    \\
Castilla                                 & 1.05    & 1.13    & 1.26     & 1.18      & 1.14                &  & 1.15    \\
Chopera                                  & 1.07    & 1.06    &          &           &                     &  & 1.06    \\
Chueca-Justicia                          & 0.73    & 0.87    & 1.00     & 0.91      & 0.87                &  & 0.87    \\
Ciudad Jardin                            & 0.95    & 1.04    & 0.90     & 0.81      & 0.81                &  & 0.90    \\
Ciudad Universitaria                     & 0.83    & 0.95    & 0.93     & 0.98      & 0.98                &  & 0.93    \\
Colina                                   & 0.48    & 0.50    &          &           &                     &  & 0.49    \\
Comillas                                 & 1.19    & 1.31    &          &           &                     &  & 1.25    \\
Concepcion                               & 0.87    & 1.01    &          &           &                     &  & 0.94    \\
Conde Orgaz-Piovera                      & 0.83    & 0.87    & 0.92     & 0.92      & 0.93                &  & 0.89    \\
Costillares                              & 0.86    & 0.87    & 1.12     & 1.17      & 1.21                &  & 1.05    \\
Cuatro Caminos                           & 0.99    & 1.23    & 1.21     & 1.10      & 1.10                &  & 1.13    \\
Cuzco-Castillejos                        & 0.60    & 0.66    & 0.77     & 0.88      & 0.88                &  & 0.76    \\
Delicias                                 & 1.23    & 1.28    & 1.20     &           &                     &  & 1.24    \\
El Burgo                                 & 1.03    & 1.08    & 1.18     & 1.10      & 1.17                &  & 1.11    \\
El Canaveral-Los Berrocales              & 1.26    & 1.14    &          &           &                     &  & 1.20    \\
El Cano-Maracaibo                        & 2.76    & 2.76    &          &           &                     &  & 2.76    \\
El Cantizal                              & 0.65    & 0.65    & 0.67     &           &                     &  & 0.66    \\
El Monte                                 & 1.00    & 1.00    & 1.00     & 1.00      & 1.00                &  & 1.00    \\
El Pinar-Punta Galea                     & 0.50    & 0.50    & 0.50     & 0.50      & 0.50                &  & 0.50    \\
El Plantio                               & 1.58    & 1.58    & 1.58     & 1.58      & 1.58                &  & 1.58    \\
ElViso                                   & 0.51    & 0.55    & 0.62     & 0.68      & 0.68                &  & 0.61    \\
EnsanchedeVallecas-LaGavia               & 1.21    & 1.15    & 1.01     &           &                     &  & 1.13    \\
Entrevias                                & 1.46    &         &          &           &                     &  & 1.46    \\
Estrella                                 & 0.86    & 0.80    & 1.00     & 1.16      & 1.16                &  & 1.00    \\
Europolis                                & 1.02    & 1.02    & 0.89     & 0.84      & 0.84                &  & 0.92    \\
Fontarron                                & 1.97    & 1.60    &          &           &                     &  & 1.78    \\
Fuente del Berro                         & 0.94    & 0.99    & 1.01     & 0.85      & 0.85                &  & 0.93    \\
Fuentelarreina                           & 0.45    & 0.45    &          &           &                     &  & 0.45    \\
Gaztambide                               & 0.85    & 0.89    & 0.91     & 0.85      & 0.93                &  & 0.89    \\
Goya                                     & 0.77    & 0.89    & 0.85     & 0.84      & 0.93                &  & 0.86    \\
Guindalera                               & 0.91    & 1.04    & 0.98     & 0.90      & 0.90                &  & 0.95    \\
Huertas-Cortes                           & 0.58    & 0.66    & 0.69     & 0.67      & 0.65                &  & 0.65    \\
Ibiza                                    & 0.70    & 0.64    & 0.58     & 0.51      & 0.49                &  & 0.58    \\
Imperial                                 & 0.99    & 0.82    &          &           &                     &  & 0.91    \\
Jeronimos                                & 0.67    & 0.74    & 0.72     & 0.74      & 0.74                &  & 0.72    \\
La Cabana                                & 1.19    & 1.19    & 1.19     & 1.19      & 1.19                &  & 1.19    \\
La Canadilla                             & 0.92    &         &          &           &                     &  & 0.92    \\
La Finca                                 & 1.36    & 1.36    & 1.36     & 1.36      & 1.36                &  & 1.36    \\
La Paz                                   & 1.12    & 1.10    & 1.04     & 1.02      & 1.02                &  & 1.06    \\
Las Dehesillas-Vereda de los Estudiantes & 1.54    & 1.54    &          &           &                     &  & 1.54    \\
LasMatas-Penascales                      & 0.96    & 1.09    & 1.09     & 1.09      & 1.09                &  & 1.06    \\
LasTablas                                & 1.00    & 1.06    & 1.03     & 1.33      & 1.22                &  & 1.13   
\end{tabular}
\end{table}

\begin{table}[] \small \renewcommand{\arraystretch}{0.3}
\begin{tabular}{llllllll}
                              & \multicolumn{5}{c}{Return Index per size}                      &  &         \\
Neighborhood                  & (30-60) & (60-90) & (90-120) & (120-150) & (\textgreater{}150) &  & Average \\
Lavapies-Embajadores          & 0.91    & 0.91    & 0.93     & 1.21      & 1.21                &  & 1.03    \\
Legazpi                       & 1.16    & 1.08    & 1.30     &           &                     &  & 1.18    \\
Lista                         & 0.90    & 0.95    & 1.05     & 1.12      & 1.14                &  & 1.03    \\
Los Angeles                   & 2.19    & 2.49    &          &           &                     &  & 2.34    \\
Los Rosales                   & 1.02    &         &          &           &                     &  & 1.02    \\
Lucero                        & 1.53    & 1.16    & 0.99     &           &                     &  & 1.23    \\
Malasana-Universidad          & 0.94    & 1.09    & 1.12     & 0.81      & 0.79                &  & 0.95    \\
Marazuela-El Torreon          & 0.62    & 0.73    & 1.16     & 1.16      & 1.16                &  & 0.96    \\
Marroquina                    & 1.18    & 1.18    &          &           &                     &  & 1.18    \\
Media Legua                   & 1.76    & 1.62    &          &           &                     &  & 1.69    \\
Mirasierra                    & 0.55    & 0.60    & 0.73     & 0.78      & 0.78                &  & 0.69    \\
Molino de la Hoz              & 0.89    & 1.14    & 1.14     & 1.14      & 1.14                &  & 1.09    \\
Montealina                    & 1.07    & 1.07    & 1.07     & 1.07      & 1.07                &  & 1.07    \\
Montecarmelo                  & 1.36    & 1.36    &          &           &                     &  & 1.36    \\
Montecillo-Pinar de las Rozas & 0.60    & 0.66    & 0.70     & 1.09      & 1.09                &  & 0.83    \\
Moscardo                      & 1.69    &         &          &           &                     &  & 1.69    \\
Nueva Espana                  & 0.50    & 0.55    & 0.69     & 0.67      & 0.66                &  & 0.62    \\
Nuevos Ministerios-Rios Rosas & 0.84    & 0.94    & 0.96     & 0.84      & 0.85                &  & 0.89    \\
Numancia                      & 1.84    & 1.29    &          &           &                     &  & 1.57    \\
Opanel                        & 2.93    &         &          &           &                     &  & 2.93    \\
Orcasitas                     & 2.07    &         &          &           &                     &  & 2.07    \\
Pacifico                      & 1.16    & 1.31    & 1.30     &           &                     &  & 1.26    \\
Palacio                       & 0.78    & 0.62    & 0.53     & 0.58      & 0.54                &  & 0.61    \\
Palomas                       & 0.93    & 0.88    & 1.12     & 1.20      & 1.20                &  & 1.07    \\
Palomeras Bajas               & 1.93    &         &          &           &                     &  & 1.93    \\
Palomeras sureste             & 1.22    & 1.28    &          &           &                     &  & 1.25    \\
Palos de Moguer               & 1.12    & 1.11    &          &           &                     &  & 1.12    \\
Parque Lisboa-LaPaz           & 1.39    & 1.39    &          &           &                     &  & 1.39    \\
Parque Mayor                  & 1.20    &         &          &           &                     &  & 1.20    \\
Parque Ondarreta-Urtinsa      & 0.82    &         &          &           &                     &  & 0.82    \\
Pavones                       & 1.46    & 1.46    &          &           &                     &  & 1.46    \\
Penagrande                    & 0.89    & 1.00    & 0.98     & 1.22      & 1.22                &  & 1.06    \\
Pilar                         & 1.60    & 1.33    & 1.25     & 1.15      & 1.15                &  & 1.30    \\
Pinar del Rey                 & 1.07    & 0.91    & 0.84     &           &                     &  & 0.94    \\
Portazgo                      & 2.44    &         &          &           &                     &  & 2.44    \\
Prado de Santo Domingo        & 0.84    & 0.62    &          &           &                     &  & 0.73    \\
Prado de Somosaguas           & 0.99    & 0.99    & 0.99     & 1.00      & 1.00                &  & 0.99    \\
Pradolongo                    & 1.53    &         &          &           &                     &  & 1.53    \\
Prosperidad                   & 0.87    & 0.90    & 0.77     &           &                     &  & 0.84    \\
Pueblo Nuevo                  & 1.53    & 1.26    & 1.10     &           &                     &  & 1.30    \\
Puerta Bonita                 & 1.11    &         &          &           &                     &  & 1.11    \\
Puerta del Angel              & 1.68    & 1.79    & 1.61     & 1.66      & 1.66                &  & 1.68    \\
Quintana                      & 1.12    & 1.15    & 1.19     &           &                     &  & 1.15    \\
Recoletos                     & 0.52    & 0.52    & 0.57     & 0.59      & 0.59                &  & 0.56    \\
Rejas                         & 1.04    & 0.99    & 0.85     & 0.79      & 0.79                &  & 0.89    \\
Rosas                         & 0.97    & 0.99    &          &           &                     &  & 0.98    \\
Salvador                      & 1.63    & 1.54    & 1.89     & 1.34      & 1.34                &  & 1.55    \\
San Andres                    & 1.63    &         &          &           &                     &  & 1.63    \\
Sanchinarro                   & 0.99    & 1.06    & 1.07     & 1.09      & 1.09                &  & 1.06    \\
San Diego                     & 1.99    &         &          &           &                     &  & 1.99    \\
San Fermin                    & 1.17    & 1.33    &          &           &                     &  & 1.25    \\
San Isidro                    & 1.17    & 1.72    & 1.43     & 1.43      &                     &  & 1.44    \\
San Juan Bautista             & 1.24    & 1.02    & 0.92     & 0.87      & 0.78                &  & 0.97    \\
San Pascual                   & 0.63    & 0.70    & 0.66     & 1.08      & 1.08                &  & 0.83    \\
Santa Eugenia                 & 1.38    & 1.38    &          &           &                     &  & 1.38    \\
Simancas                      & 1.23    & 1.00    & 0.80     &           &                     &  & 1.01    \\
Sol                           & 0.96    & 1.00    & 1.35     & 1.05      & 1.05                &  & 1.08    \\
Somosaguas                    & 1.07    & 1.07    & 1.07     & 1.03      & 1.03                &  & 1.05    \\
Timon                         & 0.70    & 0.81    & 0.61     &           &                     &  & 0.71    \\
Trafalgar                     & 0.90    & 0.89    & 0.97     & 0.91      & 0.90                &  & 0.91    \\
Tres Olivos-Valverde          & 0.77    & 1.42    & 1.42     & 1.42      & 1.42                &  & 1.29    \\
Valdeacederas                 & 1.43    & 1.59    &          &           &                     &  & 1.51    \\
Valdebebas-Valdefuentes       & 1.19    & 1.19    & 1.23     & 1.37      & 1.37                &  & 1.27    \\
Valdemarin                    & 0.62    & 0.63    & 0.67     & 0.68      & 0.66                &  & 0.65    \\
Valdezarza                    & 1.24    & 0.93    & 0.73     & 1.01      & 1.01                &  & 0.98    \\
Vallehermoso                  & 0.80    & 0.90    & 0.91     & 1.00      & 1.03                &  & 0.93    \\
Ventas                        & 1.83    & 1.53    &          &           &                     &  & 1.68    \\
Ventilla-Almenara             & 1.01    & 0.91    & 1.08     & 1.00      &                     &  & 1.00    \\
Virgen del Cortijo-Manoteras  & 1.06    & 1.12    & 0.94     & 0.79      & 0.79                &  & 0.94    \\
Vista Alegre                  & 1.58    & 0.63    &          &           &                     &  & 1.11    \\
Zarzaquemada                  & 1.69    &         &          &           &                     &  & 1.69    \\
Zona Avenida Europa           & 0.99    & 0.99    & 1.02     & 1.01      & 1.01                &  & 1.00    \\
Zona Estacion                 & 0.64    & 0.74    & 0.78     & 0.95      & 1.05                &  & 0.83    \\
Zonanorte                     & 0.91    & 0.91    & 0.99     & 1.05      & 1.05                &  & 0.98    \\
Zona Pueblo                   & 0.95    & 0.88    & 1.00     & 0.85      & 0.85                &  & 0.91   
\end{tabular}
\end{table}

\end{document}